\documentclass[preprint,aps,shownopacs,floats,superscriptaddress,nofootinbib]{revtex4}
\usepackage{psfig}
\usepackage[dvips]{graphicx,psfrag}
\usepackage{epsfig}
\newcommand{\lsim}{\buildrel<\over{_\sim}}
\newcommand{\gsim}{\buildrel>\over{_\sim}}
\newcommand{\befig}{\begin{figure}}
\newcommand{\efig}{\end{figure}}
\newcommand{\betab}{\begin{table}}
\newcommand{\etab}{\end{table}}
\newcommand{\barray}{\begin{array}}
\newcommand{\earray}{\end{array}}
\newcommand{\be}{\begin{equation}}
\newcommand{\ee}{\end{equation}}
\newcommand{\bea}{\begin{eqnarray}}
\newcommand{\eea}{\end{eqnarray}}
\newcommand{\benn}{\begin{displaymath}}
\newcommand{\eenn}{\end{displaymath}}
\newcommand{\beann}{\begin{eqnarray*}}
\newcommand{\eeann}{\end{eqnarray*}}
\newcommand{\Order}{{\cal O}}   
\newcommand{\MeV}{\mbox{MeV}}
\newcommand{\GeV}{\mbox{GeV}}

\newcommand{\alphaEM}{\alpha}
\newcommand{\MPl}{\mathrm{M}_{\mathrm{Pl}}}

\newcommand{\stau}{{\widetilde \tau}}
\newcommand{\gravitino}{{\widetilde{G}}}
\newcommand{\axino}{{\widetilde a}}
\newcommand{\Bino}{{\widetilde B}}
\newcommand{\mgravitino}{m_{\widetilde{G}}}
\begin{document}
%
%
\preprint{CERN-PH-TH/2005-005}
\preprint{DESY 05-006}
\vspace*{0.5cm}
%
%
\title{Signatures of Axinos and Gravitinos at Colliders}
%
%
\author{A.~Brandenburg}
\affiliation{DESY Theory Group, Notkestrasse 85, D--22603 Hamburg, Germany}
\author{L.~Covi}
\affiliation{CERN Theoretical Physics Division, CH--1211 Geneva 23, Switzerland}
\author{K.~Hamaguchi}
\affiliation{DESY Theory Group, Notkestrasse 85, D--22603 Hamburg, Germany}
\author{L.~Roszkowski}
\affiliation{Department of Physics and Astronomy, University of
  Sheffield, Sheffield S3 7RH, England \vspace*{0.5cm}} 
\author{F.~D.~Steffen}
\affiliation{DESY Theory Group, Notkestrasse 85, D--22603 Hamburg, Germany}
%
%
%
%
%
\begin{abstract}
  The axino and the gravitino are well-motivated candidates for the
  lightest supersymmetric particle (LSP) and also for cold dark matter
  in the Universe. Assuming that a charged slepton is the
  next-to-lightest supersymmetric particle (NLSP), we show how the
  NLSP decays can be used to probe
  the axino LSP scenario in hadronic axion models as well as
  the gravitino LSP scenario at the Large Hadron Collider and the
  International Linear Collider. We show how one can identify
  experimentally the scenario realized in nature.  In the case of the
  axino LSP, the NLSP decays will allow one to estimate the value of
  the axino mass and the Peccei--Quinn scale.
\end{abstract}
%
%
%
\maketitle
%
%
\section{Introduction}

In supersymmetric extensions of the Standard Model with unbroken
R-parity~\cite{Nilles:1983ge+X}, the lightest supersymmetric particle
(LSP) is stable and plays an important role in both collider
phenomenology and cosmology.  The most popular LSP candidate is the
lightest neutralino, which appears already in the Minimal
Supersymmetric Standard Model (MSSM). Here we consider two
well-motivated alternative LSP candidates, which are not part of the
spectrum of the MSSM: the axino and the gravitino. In particular,
either of them could provide the right amount of cold dark matter in
the Universe if heavier than about 1~MeV
(see~\cite{Covi:1999ty+X,Brandenburg:2004du+X}
and~\cite{Moroi:1993mb,Bolz:1998ek+X,Fujii:2002fv+X,Feng:2003xh+X,Roszkowski:2004jd},
respectively, and references therein).

The axino~\cite{Nilles:1981py+X,Tamvakis:1982mw,Kim:1983ia} appears (as
the spin-1/2 superpartner of the axion) when extending the MSSM with
the Peccei--Quinn mechanism~\cite{Peccei:1977hh+X} in order to solve
the strong CP problem. Depending on the model and the supersymmetry
(SUSY) breaking scheme, the mass of the axino can range between the eV
and the GeV
scale~\cite{Tamvakis:1982mw,Nieves:1985fq+X,Rajagopal:1990yx,Goto:1991gq+X}.

The gravitino appears (as the spin-3/2 superpartner of the graviton)
once SUSY is promoted from a global to a local symmetry leading to
supergravity (SUGRA)~\cite{Wess:1992cp}. The mass of the gravitino
depends strongly on the SUSY-breaking scheme and can range from the eV
scale to scales beyond the TeV
region~\cite{Nilles:1983ge+X,Dine:1994vc+X,Randall:1998uk+X}. 
In particular, in gauge-mediated SUSY breaking
schemes~\cite{Dine:1994vc+X}, the gravitino mass is typically less than 100~MeV, 
while in gravity-mediated schemes~\cite{Nilles:1983ge+X} 
it is expected to be in the GeV to TeV range.
 
Both the axino and the gravitino are singlets with respect to the
gauge groups of the Standard Model. Both interact extremely weakly as
their interactions are suppressed by the Peccei--Quinn
scale~\cite{Sikivie:1999sy,Eidelman:2004wy} $f_a\gsim 5\times
10^9\,\GeV$ and the (reduced) Planck scale~\cite{Eidelman:2004wy}
$\MPl=2.4\times 10^{18}\,\GeV$, respectively. Therefore, in both the
axino LSP and the gravitino LSP cases, the next-to-lightest
supersymmetric particle (NLSP) typically has a long lifetime.
For example, for axino cold dark matter, an NLSP with a mass of 100~GeV has a
lifetime of $\Order$(1~sec). For gravitino cold dark matter, this lifetime is of
$\Order$(1~sec) for a gravitino mass of 10~MeV and of
$\Order(10^6~\mbox{sec})$ for a gravitino mass of 10~GeV.
Late NLSP decays can spoil successful predictions of primordial
nucleosynthesis and can distort the CMB blackbody spectrum.
Constraints are obtained in order to avoid the corresponding (rather
mild) axino problem or the more severe and better-known gravitino
problem. In the axino LSP case, either a neutralino or a slepton could
be the NLSP~\cite{Covi:2004rb}. In the gravitino LSP case, these
constraints strongly disfavour a bino-dominated neutralino NLSP, while
a slepton NLSP remains
allowed~\cite{Fujii:2003nr+X,Roszkowski:2004jd}.

Because of their extremely weak interactions, the direct detection of
axinos and gravitinos seems hopeless. Likewise, their direct
production at colliders is very strongly suppressed.  Instead, one
expects a large sample of NLSPs from pair production or cascade decays
of heavier superparticles, provided the NLSP belongs to the MSSM
spectrum. These NLSPs will appear as quasi-stable particles, which
will eventually decay into the axino/gravitino LSP. A significant
fraction of these NLSP decays will take place outside the detector and
will thus escape detection. For the charged slepton NLSP scenario,
however, there have recently been proposals, which discuss the way
such NLSPs could be stopped and collected for an analysis of their
decays into the LSP.  It was found that up to $\Order(10^3$--$10^4)$
and $\Order(10^3$--$10^5)$ of charged NLSPs can be trapped per year at
the Large Hadron Collider (LHC) and the International Linear Collider
(ILC), respectively, by placing 1--10~kt of massive additional
material around planned collider
detectors~\cite{Hamaguchi:2004df,Feng:2004yi}.

In this Letter we assume that the NLSP is a charged slepton.  In
Sec.~\ref{Sec:AxinoLSP} we investigate the NLSP decays in the axino
LSP scenario. These decays were previously considered
in~\cite{Covi:2004rb}. We show that the NLSP decays can be used to
estimate the axino mass and to probe the Peccei--Quinn sector. In
particular, we obtain a new method to measure the Peccei--Quinn scale
$f_a$ at future colliders.

In Sec.~\ref{Sec:GravitinoLSP} we consider the corresponding NLSP
decays in the gravitino LSP scenario. These decays were already
studied in~\cite{Buchmuller:2004rq}. It was shown that the measurement
of the NLSP lifetime can probe the gravitino mass and can lead to a
new (microscopic) determination of the Planck scale with an
independent kinematical reconstruction of the gravitino mass.
Moreover, it was demonstrated that slepton NLSP decays into the
corresponding lepton, the gravitino, and the photon can be used to
reveal the peculiar couplings and possibly even the spin of the
gravitino. In Ref.~\cite{Buchmuller:2004rq} the limit of an infinite
neutralino mass was used. Here we generalize the result obtained
therein for the three-body decay by taking into account finite values
of the neutralino mass.

A question arises as to whether one can distinguish between the axino
LSP and the gravitino LSP scenarios at colliders.  From the NLSP
lifetime alone, such a distinction will be difficult, in particular if
the mass of the LSP cannot be determined.  Thus, an analysis of the
three-body decay of the charged NLSP slepton into the corresponding
lepton, the LSP, and a photon will be essential. With a measurement of
the polarizations of the final-state lepton and photon, the
determination of the spin of the LSP should be
possible~\cite{Buchmuller:2004rq} and would allow us to decide clearly
between the spin-1/2 axino and the spin-3/2 gravitino.  The spin
measurement, however, will be very difficult.  In
Sec.~\ref{Sec:AxinovsGravitino} we present more feasible methods to
distinguish between the axino LSP and the gravitino LSP scenarios,
which are also based on the analysis of the three-body NLSP decay with
a lepton and a photon in the final state.

Let us comment on the mass hierarchy of the relevant particles.  There
are six possible orderings in the hierarchy of the axino mass
$m_{\axino}$, the gravitino mass $\mgravitino$, and the mass of the
lightest ordinary supersymmetric particle (LOSP) $m_\mathrm{LOSP}$.
Here the LOSP is the lightest charged slepton. The cases relevant in
this Letter are (i)~$m_{\axino} < m_\mathrm{LOSP} < \mgravitino$,
(ii)~$\mgravitino < m_\mathrm{LOSP} < m_{\axino}$, (iii)~$m_{\axino} <
\mgravitino < m_\mathrm{LOSP}$, and (iv)~$\mgravitino < m_{\axino} <
m_\mathrm{LOSP}$. In cases (iii) and (iv), the LOSP has two distinct
decay channels, one into the axino and the other into the gravitino.
However, unless the decay rates into the axino and the gravitino are
(accidentally) comparable, the phenomenology of the LOSP decay in the
cases (iii) and (iv) can essentially be reduced to the cases~(i)
or~(ii), although not necessarily respectively, as will be discussed
in Sec.~\ref{Sec:AxinovsGravitino}.  We will thus concentrate on the
cases (i) and (ii) and call the LOSP the NLSP.

\section{Axino LSP Scenario}
\label{Sec:AxinoLSP}

In this section we consider the axino LSP scenario. The relevant
interactions of the axino are discussed. The rates of the two-body
and three-body decays of the charged slepton NLSP are given. We
demonstrate that these decays can be used to estimate the
Peccei--Quinn scale and the axino mass.

To be specific, we focus on the case where the lighter stau $\stau$ is
the NLSP. In general, the stau is a linear combination of $\stau_{\mathrm R}$
and $\stau_{\mathrm L}$, which are the superpartners of the right-handed and
left-handed tau lepton, respectively:
$\stau=\cos\theta_\tau\stau_{\mathrm R}+\sin\theta_\tau\stau_{\mathrm L}$. For simplicity,
we concentrate on a pure `right-handed' stau $\stau_{\mathrm R}$, which is a
good approximation at least for small $\tan\beta$. Then, the
neutralino--stau coupling is dominated by the bino coupling. In
addition, we assume for simplicity that the lightest neutralino is a
pure bino.

\subsection{Axino Interactions}
\label{Sec:AxinoInteractions}

Let us first discuss how the axino couples to the stau.
Concentrating on hadronic, or KSVZ, axion models~\cite{Kim:1979if+X}
in a SUSY setting, the coupling of the axino to the bino and the
photon/Z-boson at scales below the Peccei--Quinn scale $f_a$ is given
effectively by the Lagrangian~\cite{Covi:1999ty+X}
\begin{eqnarray}
  {\cal L}_{\axino}&=& i\,\frac{\alpha_{\mathrm{Y}} C_{\rm aYY}}{16\pi f_a}\,
  {\overline {\tilde a}}\,\gamma_5\,[\gamma_\mu,\gamma_\nu]\,
  \Bino\,
  \left(
  \cos\theta_W F_{\mu\nu}-\sin\theta_W Z_{\mu\nu}
  \right) ,
  \label{Eq:AxinoInteractions}
\end{eqnarray}
where $\theta_W$ is the weak mixing angle,
$\alpha_{\mathrm{Y}}=\alphaEM/\cos^2\theta_W$ with the fine structure constant
$\alpha$, and $F_{\mu\nu}$ and $Z_{\mu\nu}$ are the field strength
tensors of the photon and Z-boson, respectively.  The interaction
Lagrangian~(\ref{Eq:AxinoInteractions}) is obtained by integrating out
the heavy (s)quarks introduced in supersymmetric KSVZ axion models.
Indeed, the KSVZ axino couples directly only to these additional heavy
(s)quarks.  Thus, the above coupling depends, for example, on the
hypercharge of these heavy (s)quarks, which we assume to be non-zero.
The model dependence related to the Peccei--Quinn sector
is expressed in terms of the factor $C_{\rm aYY}\simeq
\mathcal{O}(1)$. As the MSSM fields do not carry Peccei--Quinn
charges, the axino couples to the stau only indirectly, via the
exchange of intermediate gauge bosons and gauginos.

In the alternative DFSZ axion models~\cite{Dine:1981rt+X}, once
supersymmetrized, the mixing of the axino with the MSSM neutralinos
can be non-negligible and other couplings between the axino and the
MSSM fields will arise. Here, however, we focus on the KSVZ-type models.

\vspace*{1cm}

\subsection{The Two-Body Decay \boldmath{$\stau \to \tau + \axino$} }
\label{Sec:Axino2Body}

We now consider the two-body decay $\stau\to\tau+\axino$ in the
framework described above. We neglect the tau mass for simplicity.
With the effective vertex~(\ref{Eq:AxinoInteractions}), i.e.\ 
with the heavy KSVZ (s)quarks integrated out, this two-body decay
occurs at the one-loop level. The corresponding Feynman diagrams are
shown in Fig.~\ref{Fig:Axino2Body}, where the effective vertex is
indicated by a thick dot.
\befig
\centerline{\epsfig{file=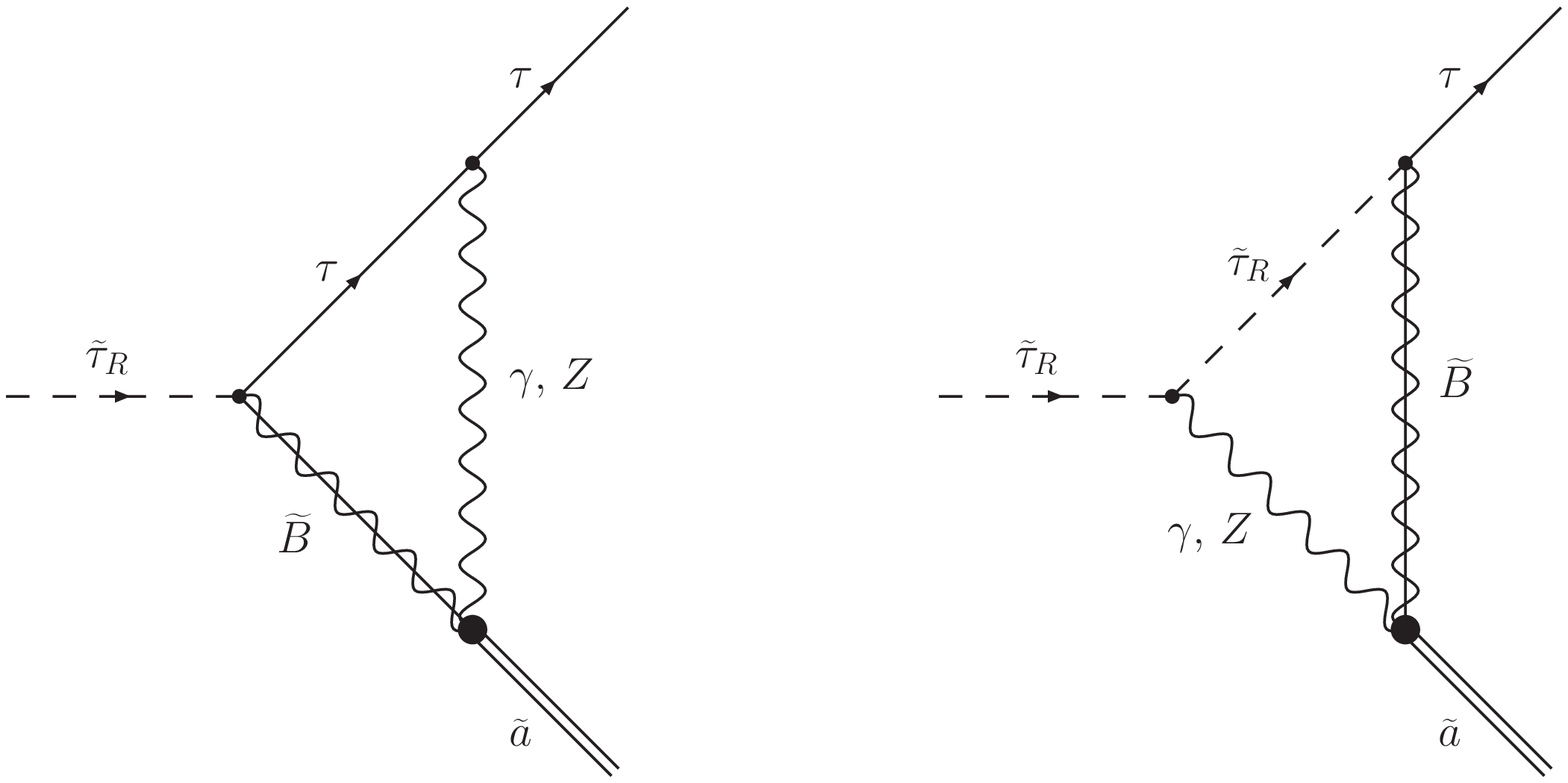,width=14cm}}
\caption{
  The dominant contributions to the two-body NLSP decay $\stau_{\mathrm R} \to
  \tau +\axino$.}
\label{Fig:Axino2Body}
\efig
Using the method described in~\cite{Covi:2002vw}, we obtain the
following estimate for the decay rate:\footnote{We correct the factor
  of $(1/16)(1+\tan^2\theta_W)^2/(1-\tan^2\theta_W)^2$, which is
  missing in Eq.~(3.12) of Ref.~\cite{Covi:2004rb}.}
\bea
  \Gamma(\stau_{\mathrm R} \to\tau\,\axino)
  &=&
  \frac{9\,\alpha^4\,C_{\rm aYY}^2}{512\pi^5\cos^8\theta_W}\,\,
  \frac{m_{\Bino}^2}{f_a^2}\,\,
  \frac{(m_{\stau}^2 - m_{\axino}^2)^2}{m_{\stau}^3}\,\,
  \xi^2\,
  \log^2\left(\frac{f_a}{m}\right)
\label{Eq:Axino2BodyI}
  \\
  &\simeq& \xi^2\,(25~\mathrm{sec})^{-1}
  C_{\rm aYY}^2
  \left(1-\frac{m_\axino^2}{m_\stau^2}\right)
  \left(\frac{m_{\stau}}{100\,\GeV}\right)
  \left(\frac{10^{11}\,\GeV}{f_a}\right)^2
  \left(\frac{m_{\tilde{B}}}{100\,\GeV}\right)^2 ,
\label{Eq:Axino2BodyII}
\eea
where $m_{\Bino}$ is the mass of the bino and $m_\stau$ is the mass of
the stau NLSP, i.e.\ $m_\axino < m_\stau < m_{\Bino}$. As explained
below, there is an uncertainty associated with the method used to
derive the decay rate~(\ref{Eq:Axino2BodyI}). We absorb this uncertainty
into the mass scale $m\simeq m_{\stau,\Bino} \simeq \Order(100\,\GeV)$
and into the factor $\xi\simeq\Order(1)$ in the first line. We used
$\log\left(f_a/m\right)\simeq 20.7$ to get from the first to the
second line.

Here a technical comment on the loop integral is in order. If one
naively integrates over the internal momentum in the diagrams with the
effective vertex~---~see~Fig.~\ref{Fig:Axino2Body}~---~one encounters
logarithmic divergencies. This is because the effective
vertex~(\ref{Eq:AxinoInteractions}) is applicable only if the momentum
is smaller than the heavy (s)quark masses, whereas the momentum in the
loop goes beyond that scale. In a rigorous treatment, one has to
specify the origin of the effective vertex, i.e.\ the Peccei--Quinn
sector, and to calculate the two-loop integrals with heavy (s)quarks
in the additional loop. Such a two-loop computation leads to a finite
result~\cite{HSSW:2005}. Here, instead, we have regulated the
logarithmic divergencies with the cut-off $f_a$ and kept only the
dominant contribution. The mass scale $m$ and the factor $\xi$ have
been introduced above to account for the uncertainty coming from this
cut-off procedure.

\subsection{The Three-Body Decay \boldmath{$\stau \to \tau + \gamma + \axino$} }
\label{Sec:Axino3Body}

We now turn to the three-body decay
$\stau_{\mathrm R}\to\tau+\gamma+\axino$.  We again neglect the tau mass for
simplicity. In contrast to the two-body decay considered above, the
three-body decay occurs already at tree level, once the effective
vertex given in~(\ref{Eq:AxinoInteractions}) is used.
In addition, we take into account photon radiation from the loop
diagrams of Fig.~\ref{Fig:Axino2Body}, since the additional factor of
$\alpha$ is partially compensated by the additional factor of $\log(f_a/m)$. 
As above, we keep only the dominant contribution of the loop diagrams.
The corresponding Feynman diagrams are shown in Fig.~\ref{Fig:Axino3Body}, 
\befig
\centerline{\epsfig{file=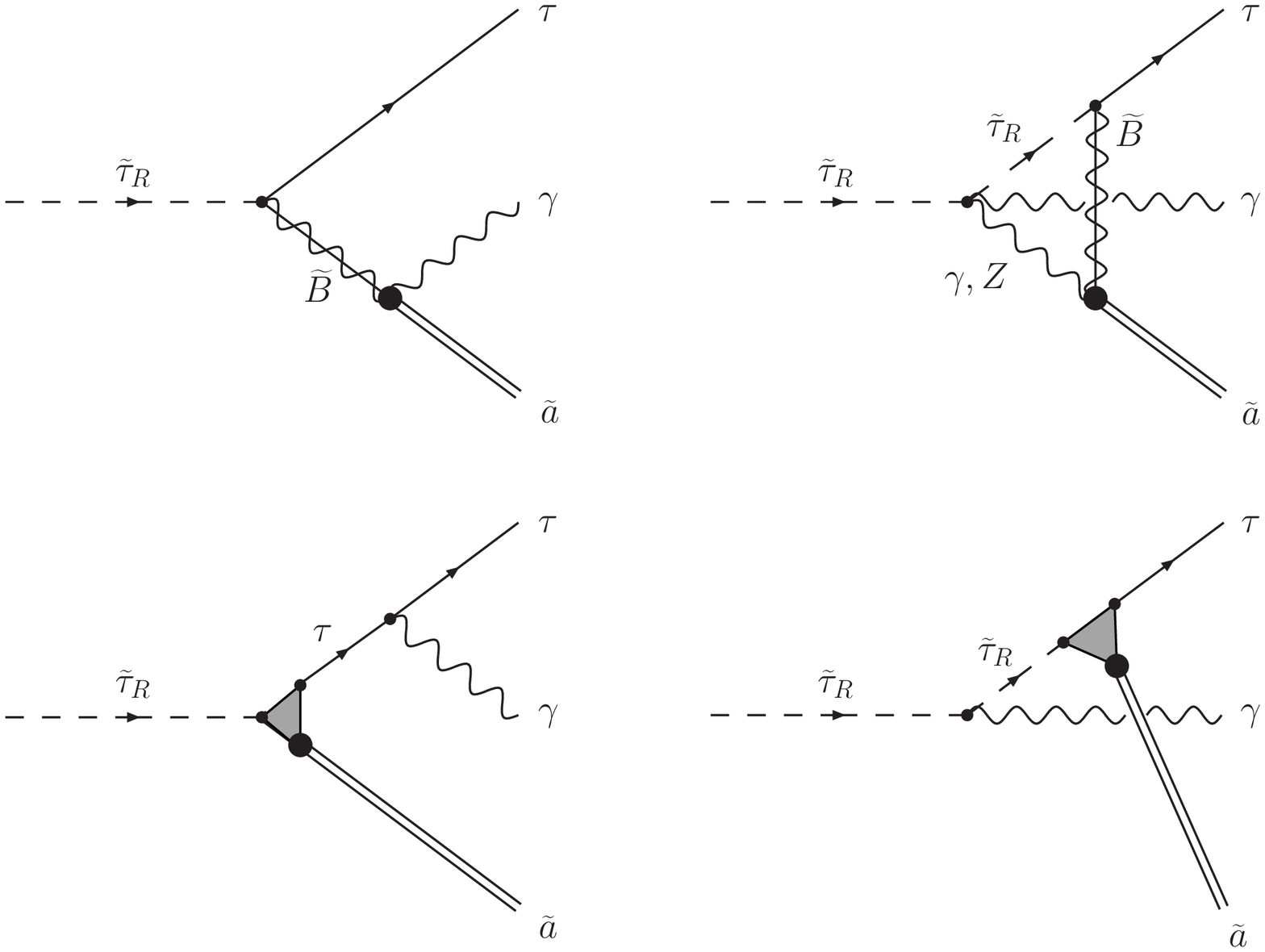,width=13.5cm}}
\bigskip
\caption{
The dominant contributions to the three-body NLSP decay 
$\stau_{\mathrm R} \to \tau + \gamma +\axino$.}
\label{Fig:Axino3Body}
\efig
where a thick dot represents the effective vertex~(\ref{Eq:AxinoInteractions}) 
and a shaded triangle the set of triangle diagrams given in Fig.~\ref{Fig:Axino2Body}.
As the photon radiation from an electrically charged particle within the loops 
leads to a subdominant contribution, these processes are not shown in Fig.~\ref{Fig:Axino3Body}. 
At each order in $\log(f_a/m)$, only the leading order in $\alpha$ is computed 
while higher-order corrections are not considered. 
In terms of the observables that seem to be most accessible, i.e.\ the
photon energy $E_{\gamma}$ and $\cos\theta$, the cosine of the opening
angle between the photon and the tau direction, the corresponding
differential decay rate reads
\be
  \frac{d^2\Gamma(\stau_{\mathrm R}\to\tau\,\gamma\,\axino)}{dx_\gamma\,d\cos\theta}
  = 
  \frac{m_{\stau}}{512\pi^3}\,
  \frac{x_\gamma(1-A_\axino-x_\gamma)}{[1-(x_\gamma/2)(1-\cos\theta)]^2}\,
  \sum_{\rm spins} |\mathcal{M}(\stau_{\mathrm R}\to\tau\,\gamma\,\axino)|^2 \ ,
\ee
where
\bea
  \sum_{\rm spins} |\mathcal{M}(\stau_{\mathrm R}\to\tau\,\gamma\,\axino)|^2
  & = &
  {\alphaEM^3\,\*C_{\rm aYY}^2\over \pi\*\cos^4\theta_W}\,\,
  \*{m_{\stau}^2\over f_a^2}\*\,\,
  F_{\rm diff}^{(\axino)}(x_\gamma,\cos\theta,A_\axino,A_\Bino)
  \ ,
\eea
with
\be
        x_{\gamma} \equiv \frac{2 E_{\gamma}}{m_{\stau}}
        \ ,
        \quad\quad
        A_\axino \equiv \frac{m_\axino^2}{m_{\stau}^2}
        \ ,
        \quad\quad
        A_\Bino \equiv \frac{m_\Bino^2}{m_{\stau}^2}
        \ ,
\label{Eq:xgammaAi}
\ee
and
\bea
  &&
  F_{\rm diff}^{(\axino)} (x_\gamma,\cos\theta,A_\axino,A_\Bino)
  =
  {x_\gamma^2 \*(1 \!-\! A_\axino \!-\! x_\gamma)\*
        [1 \!+\! \cos\theta \!+\! A_\axino\* (1 \!-\! \cos\theta)]\*
        [1 \!+\! \cos\theta \!+\! A_\Bino  \*(1 \!-\! \cos\theta)]\over
  \{x_\gamma\*(1+\cos\theta)+2\*A_\axino-A_\Bino\*[2- x_\gamma
  (1-\cos\theta)] \}^2}
\nonumber\\
  &&
  +\,\, {3\*\alphaEM\over \pi\*\cos^2\theta_W}\*\,
  \xi\*\log\left({f_a\over m}\right)
  \*\Bigg\{ {\sqrt{A_\axino\*A_\Bino}
  \*(1+\cos\theta)\*(1-A_\axino-x_\gamma)\over
  x_\gamma\*(1+\cos\theta)+2\*A_\axino-A_\Bino
  \*[2-x_\gamma\*(1-\cos\theta)]} 
\nonumber \\
  &&
  \hskip 4.5cm
  +\, {A_\Bino\*\left[(1+\cos\theta)\*(1-A_\axino)+
  A_\axino\*x_\gamma\*(1-\cos\theta)\right]\over
  x_\gamma\*(1+\cos\theta)+2\*A_\axino-A_\Bino
  \*[2-x_\gamma\*(1-\cos\theta)]}\Bigg\} 
\nonumber \\
  &&
  +\,\, {9\*\alphaEM^2\over4\*\pi^2\*\cos^4\theta_W}\*\,
  \xi^2\*\log^2\left({f_a\over m}\right)\*
  A_\Bino
  \*\Bigg\{
  {1+\cos\theta+A_\axino\*(1-\cos\theta)\over
  (1-\cos\theta)
  \*(1-A_\axino-x_\gamma)}+{2\*(1+\cos\theta)\*(1-A_\axino)
  \over x_\gamma^2\*(1-\cos\theta)}
  \Bigg\}
\ .
\eea
Hereafter, we use $\log\left(f_a/m\right)\simeq 20.7$, 
as in the previous section.

The three-body decay $\stau\to\tau+\gamma+\axino$ involves
bremsstrahlung processes (see~Fig.~\ref{Fig:Axino3Body}) and, as
already mentioned, we have neglected the tau mass.  Thus, when the
photon energy and/or the angle between the photon and the tau
direction tend to zero, there are soft and/or collinear divergences.
Consequently, the total rate of the decay
$\stau\to\tau+\gamma+\axino$ is not defined. We define the
integrated rate of the three-body decay $\stau \to \tau + \gamma
+\axino$ with a cut on the scaled photon energy, $x_\gamma >
x_\gamma^{\mathrm{cut}}$, and a cut on the cosine of the opening angle,
$\cos\theta < 1-x_\theta^{\mathrm{cut}}$:
\be
\Gamma(\stau_{\mathrm R}\to\tau\,\gamma\,\axino\,;
x_\gamma^{\mathrm{cut}},x_\theta^{\mathrm{cut}})
  \equiv
  \int^{1-A_{\tilde{a}}}_{x_\gamma^{\mathrm{cut}}}
  d x_\gamma
  \int^{1-x_\theta^{\mathrm{cut}}}_{-1}
  d \cos\theta
  \frac{d^2\Gamma(\stau_{\mathrm R}\to\tau\,\gamma\,\axino)}{dx_\gamma d\cos\theta}
  \ .
\label{Eq:Axino3BodywithCuts}
\ee
As explained in Sec.~\ref{Sec:AxinovsGravitino}, the quantity
$\Gamma(\stau_{\mathrm R}\to\tau\,\gamma\,\axino\,; x_\gamma^{\mathrm{cut}},x_\theta^{\mathrm{cut}})$ 
will be important in distinguishing between the axino LSP 
and the gravitino LSP scenarios.

\subsection{Probing the Peccei--Quinn Scale and the Axino Mass}

In the axino LSP scenario, the stau NLSP decays provide us with a new
method to probe the Peccei--Quinn scale $f_a$ at colliders. As
we will see in Sec.~\ref{Sec:BranchingRatio}, the branching ratio of
the three-body decay is small if reasonable cuts are used. Thus, we can
use the two-body decay rate~(\ref{Eq:Axino2BodyII}) to estimate the
stau lifetime, $\tau_\stau \approx 1/\Gamma(\stau\to\tau\,\axino)$.
Accordingly, the Peccei--Quinn scale $f_a$ can be estimated as
\be
        f_a^2 \simeq
        \left(\frac{\tau_\stau}{25~\mathrm{sec}}\right)\, 
        \xi^2\,C_{\rm aYY}^2 
        \left(1-\frac{m_\axino^2}{m_\stau^2}\right)
        \left(\frac{m_{\stau}}{100\,\GeV}\right)
        \left(\frac{m_{\tilde{B}}}{100\,\GeV}\right)^2
        \left(10^{11}\,\GeV\right)^2
        \ ,
\label{Eq:PQScale}
\ee
once $m_\stau$, $m_\Bino$, and the lifetime of the stau $\tau_\stau$
have been measured. The dependence on the axino mass is negligible for
$m_\axino/m_\stau\lsim 0.1$, so that $f_a$ can be determined without
knowing $m_\axino$. For larger values of $m_\axino$, the stau NLSP
decays can be used to determine the mass of the axino kinematically.
In the two-body decay $\stau\to\tau+\axino$, the axino mass can be
inferred from $E_\tau$, the energy of the emitted tau lepton:
\be
        m_\axino = \sqrt{m_\stau^2+m_\tau^2-2m_\stau E_\tau}
        \ ,
\label{Eq:AxinoMass2Body}
\ee
with an error depending on the experimental uncertainty on $m_\stau$
and $E_\tau$.

\section{Gravitino LSP Scenario}
\label{Sec:GravitinoLSP}

In this section we assume that the gravitino is the LSP and again that
the pure right-handed stau is the NLSP. The corresponding rates of
the two-body and three-body decay of the stau NLSP are given.  These
decays have already been studied in Refs.~\cite{Buchmuller:2004rq}.
Here we generalize the result obtained for the three-body decay by
taking into account finite values of the neutralino mass. For
simplicity, we assume again that the lightest neutralino is a pure
bino.

The couplings of the gravitino $\gravitino$ to the $\stau_{\mathrm R}$, $\tau$,
$\Bino$, and $\gamma$ are given by the SUGRA Lagrangian~\cite{Wess:1992cp}. 
The interactions of the gravitino are determined uniquely by local SUSY and 
the Planck scale and, in constrast to the axino case, are not model-dependent.

\subsection{The Two-Body Decay \boldmath{$\stau \to \tau + \gravitino$} }
\label{Sec:Gravitino2Body}

In the gravitino LSP scenario, the main decay mode of the stau NLSP is
the two-body decay $\stau\to\tau+\gravitino$. As there is a direct
stau--tau--gravitino coupling, this process occurs at tree level.
Neglecting the $\tau$-lepton mass $m_\tau$, one obtains the decay
rate:
\bea
  \Gamma(\stau_{\mathrm R}\to\tau\,\gravitino)
  &=&
  \frac{m_{\stau}^5}{48\pi\,\mgravitino^2\,\MPl^2}
  \left(
  1 - \frac{\mgravitino^2}{m_{\stau}^2}
  \right)^4
\label{Eq:Gravitino2BodyI}\\
  &=&
  (5.89~\mathrm{sec})^{-1}
  \left(\frac{m_{\stau}}{100~\mathrm{GeV}}\right)^5
  \left(\frac{10~\mathrm{MeV}}{\mgravitino}\right)^2
  \left(
  1 - \frac{\mgravitino^2}{m_{\stau}^2}
  \right)^4
  \ .
\label{Eq:Gravitino2BodyII}
\eea
In order to get from the first to the second line, we have used the
value of the reduced Planck mass
$\MPl = (8\pi\,G_{\rm N})^{-1/2} = 2.435\times 10^{18}\,\GeV$ 
as obtained from macroscopic measurements of Newton's
constant~\cite{Eidelman:2004wy}
$G_{\rm N} = 6.709\times 10^{-39}\,\GeV^{-2}$.
Thus, the gravitino mass can be determined once the stau NLSP lifetime
governed by~(12) and $m_\stau$ are measured.  As pointed out in
Refs.~\cite{Buchmuller:2004rq}, expression~(\ref{Eq:Gravitino2BodyI})
can also be used the other way around, i.e.\ for a microscopic
determination of the Planck scale once the masses of the gravitino and
the stau are measured kinematically.
Note the strong dependence on $\mgravitino$ and $m_\stau$. 
In the axino LSP scenario, the corresponding
rate~(\ref{Eq:Axino2BodyI}) becomes independent of the axino mass for
$m_{\axino}/m_{\stau} \lsim 0.1$, so that the Peccei--Quinn scale can
be determined even if $m_{\axino}$ is too small to be inferred
kinematically.

\vspace*{-0.2cm}

\subsection{The Three-Body Decay \boldmath{$\stau \to \tau + \gamma + \gravitino$} }
\label{Sec:Gravitino3Body}

Let us now turn to the three-body decay $\stau_{\mathrm
  R}\to\tau+\gamma+\gravitino$. The corresponding Feynman diagrams are
shown in Fig.~\ref{Fig:Gravitino3Body}.
\befig
\centerline{
\epsfig{file=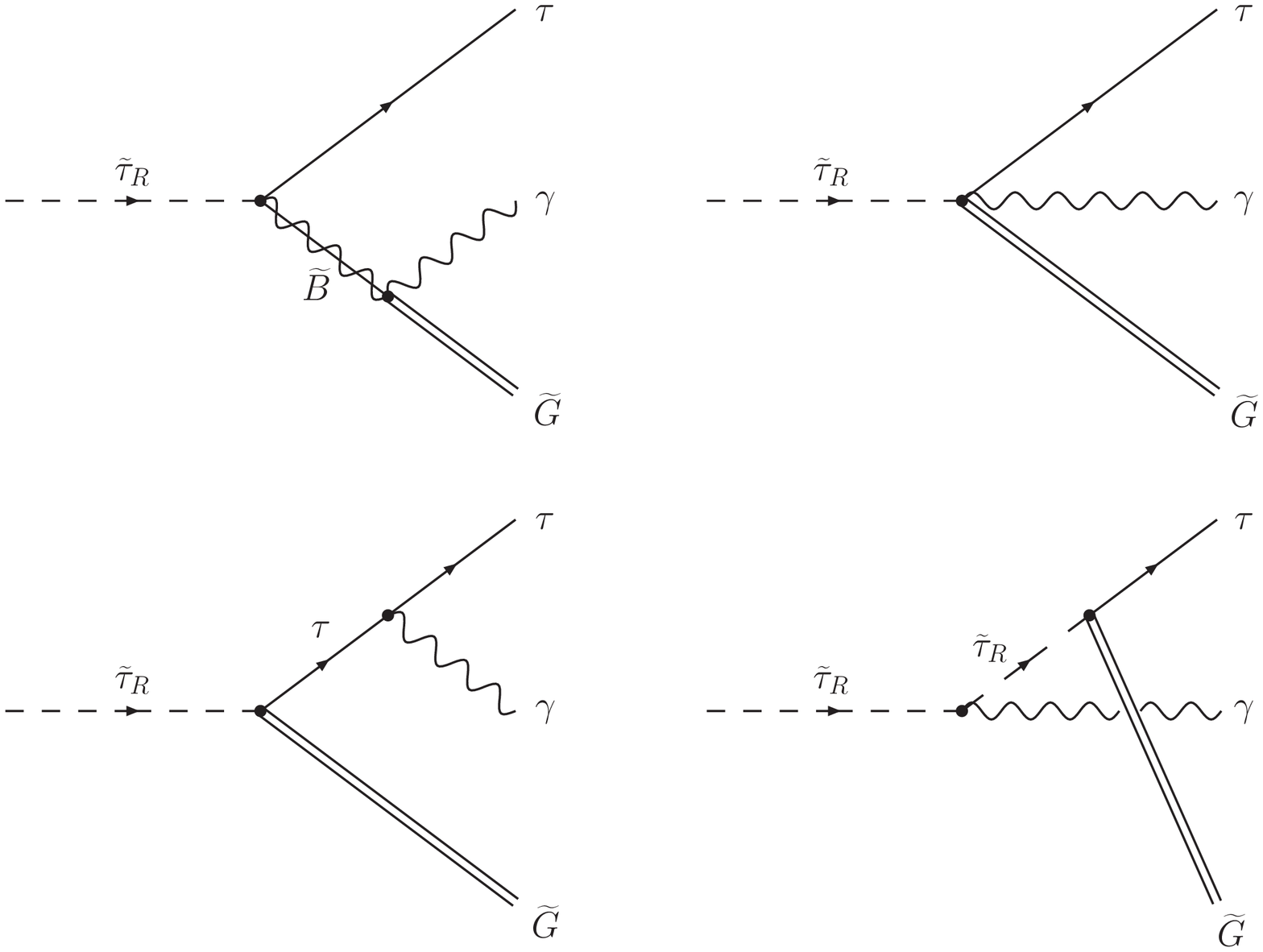,width=13.5cm}
}
\caption{
The three-body NLSP decay $\stau_{\mathrm R} \to \tau + \gamma + \gravitino$.}
\label{Fig:Gravitino3Body}
\efig
We neglect again the tau mass for simplicity.
For finite bino mass, we obtain the following differential decay rate
\be
  \frac{d^2\Gamma(\stau_{\mathrm R}\to\tau\,\gamma\,\gravitino)}{dx_\gamma d\cos\theta}
  =
  \frac{m_{\tilde{\tau}}}{512\pi^3}
  \frac{x_\gamma(1-A_{\gravitino}-x_\gamma)}{[1-(x_\gamma/2)(1-\cos\theta)]^2}
  \sum_{\rm spins} |\mathcal{M}(\stau_{\mathrm R}\to\tau\,\gamma\,\gravitino)|^2,
\ee
where
\be
  \sum_{\rm spins} |\mathcal{M}(\stau_{\mathrm R}\to\tau\,\gamma\,\gravitino)|^2
  = 
  \frac{8\pi\alphaEM}{3}\,\,
  \frac{m_\stau^2}{\MPl^2\,A_{\gravitino}}\,\,
  F_{\rm diff}^{(\gravitino)}(x_\gamma,\cos\theta,A_{\gravitino},A_\Bino)
\ee
with the definitions of $x_{\gamma}$ and $A_\Bino$ given in~(\ref{Eq:xgammaAi}),
$A_\gravitino \equiv m_\gravitino^2/m_\stau^2$, and
\bea
  &&
  F_{\rm diff}^{(\gravitino)}
  (x_\gamma,\cos\theta,A_{\gravitino},A_\Bino)
  =
  -3\*A_{\gravitino}^2
  - 7\*x_{\gamma}\*A_{\gravitino}
  +{2\*(2-5\*\cos\theta)\*A_{\gravitino}
  \over
  1-\cos\theta}
  -{x_{\gamma}\*(1+\cos\theta)
  \over
  (1-\cos\theta)}
\nonumber \\ &&\quad
  -{(1+\cos\theta)\*(3+\cos\theta)
  \over
  (1-\cos\theta)^2}
  + {2\*(1-A_{\gravitino})^3\*(1+\cos\theta)
  \over
  x_{\gamma}^2\*(1-\cos\theta)}
  + {A_{\gravitino}\*(1-A_{\gravitino})^2
  \over
  1-A_{\gravitino}-x_{\gamma}}
  \nonumber \\ &&\quad
  + {(1-A_{\gravitino})^2\*(1+\cos\theta)
  \over
  (1-A_{\gravitino}-x_{\gamma})\*(1-\cos\theta)}
  - {4\*\left[1+\cos\theta+A_{\gravitino}\*(1-\cos\theta)\right]^2
  \over
  \left[2-x_{\gamma}\*(1-\cos\theta)\right]^2\*(1-\cos\theta)^2}
\nonumber \\ &&\quad
  +
{2\*\left\{3+\cos\theta\*\left[4-\cos\theta+2\*A_{\gravitino}\*(1-\cos\theta)\right]\right\}
  \*\left[1+\cos\theta+A_{\gravitino}\*(1-\cos\theta)\right]
  \over
  \left[2-x_{\gamma}\*(1-\cos\theta)\right]\*(1-\cos\theta)^2}
\nonumber \\ &&\quad
  + 2\*(1-A_{\gravitino}-x_{\gamma})
  \*\Bigg\{
  {1+x_{\gamma}-x_{\gamma}^2
     - 2\*A_{\gravitino}\*(1+3\*x_{\gamma}-2\*x_{\gamma}^2)
     + A_{\gravitino}^2\*(1+5\*x_{\gamma})
  \over
  x_{\gamma}\*(1-A_{\tilde{B}})\*(1-A_{\gravitino}-x_{\gamma})}
\nonumber  \\ &&\quad
  - {2\*\left[1+x_{\gamma}\*(2+A_{\tilde{B}})-x_{\gamma}^2
     + 2\*A_{\gravitino}\*(1-x_{\gamma})\right]
  \over
  x_{\gamma}\*\left[2-x_{\gamma}\*(1-\cos\theta)\right]}
  + {4\*(1-A_{\gravitino}-x_{\gamma})
  \over
  \left[2-x_{\gamma}\*(1-\cos\theta)\right]^2}
\nonumber \\ &&\quad
  -{\sqrt{A_{\Bino}\*A_{\gravitino}}\*
  \left[2\*(1+\cos\theta)\*(1-A_{\gravitino})+3\*x_{\gamma}\*A_{\gravitino}\*(1-\cos\theta)\right]
  \over
  x_{\gamma}\*(1+\cos\theta)+2\*(A_{\gravitino}-A_{\tilde{B}})
  +A_{\tilde{B}}\*x_{\gamma}\*(1-\cos\theta)}
\nonumber \\ &&\quad
  -
{2\*\left\{A_{\gravitino}^2\*\left[-3-6\*x_{\gamma}+A_{\tilde{B}}\*(2+x_{\gamma})\right]
  + 4\*A_{\Bino}\*A_{\gravitino}\*(1+x_{\gamma}-x_{\gamma}^2)\right\}
  \over
  x_{\gamma}\*(1-A_{\tilde{B}})
  \*\left[x_{\gamma}\*(1+\cos\theta)+2\*(A_{\gravitino}-A_{\tilde{B}})
  +A_{\tilde{B}}\*x_{\gamma}\*(1-\cos\theta)
  \right]}
\nonumber \\ &&\quad
  + {2\*A_{\tilde{B}}^2\*\left[(1-x_{\gamma})\*(1+2\*A_{\gravitino}+x_{\gamma})
  +x_{\gamma}\*A_{\tilde{B}}\right]
  \over
  x_{\gamma}\*(1-A_{\tilde{B}})
  \*\left[x_{\gamma}\*(1+\cos\theta)+2\*(A_{\gravitino}-A_{\tilde{B}})
  +A_{\tilde{B}}\*x_{\gamma}\*(1-\cos\theta)
  \right]}
\Bigg\}
\nonumber \\ &&\quad
  + (1-A_{\gravitino}-x_{\gamma})
  \*\Bigg\{
  {(-1+3\*A_{\gravitino})\*(1-A_{\gravitino})
  \over
  (1-A_{\tilde{B}})}
  +{2\*\left[2-x_{\gamma}-2\*(A_{\gravitino}-A_{\tilde{B}})\right]
  \over
  2-x_{\gamma}\*(1-\cos\theta)}
\nonumber \\ &&\quad
  - {4\*(1-A_{\gravitino}-x_{\gamma})
  \over
  \left[2-x_{\gamma}\*(1-\cos\theta)\right]^2}
  - {2\*(A_{\gravitino}-A_{\tilde{B}})
     \*\left[3\*A_{\gravitino}\*(2-2\*A_{\gravitino}-x_{\gamma})
             +A_{\tilde{B}}\*(2-2\*A_{\tilde{B}}+x_{\gamma})\right]
  \over
  (1-A_{\tilde{B}})\*\left[x_{\gamma}\*(1+\cos\theta)
        +2\*(A_{\gravitino}-A_{\tilde{B}})
        + A_{\tilde{B}}\*x_{\gamma}\*(1-\cos\theta)\right]}
\nonumber \\ &&\quad
  +
{4\*(1-A_{\gravitino}-x_{\gamma})\*(3\*A_{\gravitino}+A_{\tilde{B}})\*(A_{\gravitino}-A_{\tilde{B}})^2
  \over
  (1-A_{\tilde{B}})\*\left[x_{\gamma}\*(1+\cos\theta)
        +2\*(A_{\gravitino}-A_{\tilde{B}})
        + A_{\tilde{B}}\*x_{\gamma}\*(1-\cos\theta)\right]^2}
\Bigg\} \ .
\label{Eq:FdiffGravitino}
\eea
In the limit $m_{\Bino}\to\infty$, only the terms in the first four
lines of~(\ref{Eq:FdiffGravitino}) remain and the result given in the
appendix of the first reference in~\cite{Buchmuller:2004rq} is
obtained. For finite values of the bino mass, the diagram with the
bino propagator in Fig.~\ref{Fig:Gravitino3Body} has to be taken into
account, which then leads to our more general result.

As in the axino case, the total rate of the three-body decay $\stau
\to \tau + \gamma +\gravitino$ is not defined. We thus introduce again
the integrated rate with a cut on the scaled photon energy, $x_\gamma
> x_\gamma^{\mathrm{cut}}$, and a cut on the cosine of the opening angle,
$\cos\theta < 1-x_\theta^{\mathrm{cut}}$,
\bea
\Gamma(\stau_{\mathrm R}\to\tau\,\gamma\,\gravitino\,;
x_\gamma^{\mathrm{cut}},x_\theta^{\mathrm{cut}})
  &=&
  \int^{1-A_{\tilde{G}}}_{x_\gamma^{\mathrm{cut}}}
  d x_\gamma
  \int^{1-x_\theta^{\mathrm{cut}}}_{-1}
  d \cos\theta
  \frac{d^2\Gamma(\stau_{\mathrm R}\to\tau\,\gamma\,\gravitino)}
    {dx_\gamma d\cos\theta}
  \ .
\eea
This quantity will be used in our comparison of collider signatures of
the axino LSP and the gravitino LSP scenarios.

\section{Axino vs.\ Gravitino}
\label{Sec:AxinovsGravitino}

In this section we show how the two-body and three-body decays of
the stau NLSP can be used to distinguish between the axino LSP
scenario and the gravitino LSP scenario at colliders. We compare the
total decay rates of the stau NLSP, the branching ratios of the
three-body decays $\stau\to\tau+\gamma+\axino/\gravitino$ with cuts
on the observables, and the differential distributions of the decay
products in the three-body decays.

\subsection{Total Decay Rates}
\label{Sec:TotalDecayRates}

Let us discuss the lifetime of the stau NLSP in the axino LSP
and in the gravitino LSP scenarios, and examine whether the lifetime
can be used to distinguish between the two. In both cases, the total
decay rate of the stau NLSP is dominated by the two-body decay,
\be
        \Gamma^{\rm{total}}_{\tilde{\tau}_R\,\to\, i\,X} 
        \simeq
        \Gamma(\stau_{\mathrm R}\to\tau\, i)
        \ ,
        \quad 
        i = \axino, \gravitino
        \ ,
\label{Eq:TotalRates}
\ee
with the rates given respectively in~(\ref{Eq:Axino2BodyII})
and~(\ref{Eq:Gravitino2BodyII}).  Thus, the order of magnitude of the
stau NLSP lifetime is (essentially) determined by $m_\stau$,
$m_\Bino$, and $f_a$ in the axino LSP scenario and by $m_\stau$ and
$\mgravitino$ in the gravitino LSP scenario. Among those parameters,
one should be able to measure the stau mass $m_\stau$ and the bino
mass $m_\Bino$ by analysing the other processes occurring in the
planned collider detectors. Indeed, we expect that these masses will
already be known when the stau NLSP decays are analysed.  To be
specific, we set these masses to $m_\stau=100\,\GeV$ and
$m_\Bino=110\,\GeV$, keeping in mind the NLSP lifetime dependencies
$\tau_\stau \propto 1/(m_\stau \,m_\Bino^2)$ for the axino LSP and
$\tau_\stau \propto 1/m_\stau^5$ for the gravitino LSP.  Then, the
order of magnitude of the stau NLSP lifetime is governed by the
Peccei--Quinn scale $f_a$ in the axino LSP scenario and by the
gravitino mass $\mgravitino$ in the gravitino LSP scenario.

In the axino LSP scenario, the stau lifetime varies from
$\Order(0.01~{\mbox{sec}})$ to $\Order(10~{\mbox{h}})$ if we change
the Peccei--Quinn scale $f_a$ from $5\times 10^9\,\GeV$ to $5\times
10^{12}\,\GeV$, as can be seen from~(\ref{Eq:Axino2BodyII}).  For the
given values of $m_\stau$ and $m_\Bino$, these values can probably be
considered as the lower and upper bounds on the stau NLSP lifetime in
the axino LSP case.

In the gravitino LSP case, the stau lifetime can vary over a much
wider range, e.g.\ from $6\times 10^{-8}\,{\rm sec}$ to 15~years by
changing the gravitino mass $\mgravitino$ from 1~keV to 50~GeV, as can
be seen from~(\ref{Eq:Gravitino2BodyII}). Therefore, both a very short
stau NLSP lifetime, $\tau_\stau \lsim$~msec, and a very long one,
$\tau_\stau \gsim$~days, will point to the gravitino LSP scenario. For
example, in gravity-mediated SUSY breaking models, the gravitino mass
is typically $(10$--$100)\,\GeV$. Then, the lifetime of the NLSP
becomes of $\Order(\mbox{years})$ and points clearly to the gravitino
LSP scenario.

On the other hand, if the observed lifetime of the stau NLSP is within
the range $\Order(0.01~{\mbox{sec}})$--$\Order(10~{\mbox{h}})$, it
will be very difficult to distinguish between the axino LSP and the
gravitino LSP scenarios from the stau NLSP lifetime alone. In this
case, the analysis of the three-body NLSP decays will be crucial to
distinguish between the two scenarios.

\subsection{Branching Ratio of the Three-Body Decay Modes}
\label{Sec:BranchingRatio}

We now consider the branching ratio of the integrated rate of the
three-body decay $\stau\to\tau+\gamma+\axino/\gravitino$ with cuts
\be
        BR(\stau_{\mathrm R}\to\tau\,\gamma\, i\,;
        x_{\gamma}^{\mathrm{cut}},x_{\theta}^{\mathrm{cut}})
        \equiv
        {\Gamma(\stau_{\mathrm R}\to\tau\,\gamma\, i\,;
          x_\gamma^{\mathrm{cut}},x_\theta^{\mathrm{cut}})
        \over
        \Gamma^{\rm{total}}_{\tilde{\tau}_R\,\to\, i\,X} }
        \ ,
        \quad 
        i = \axino, \gravitino
        \ .
\label{Eq:BranchingRatio}
\ee
In Fig.~\ref{Fig:BranchingRatio}
\befig
\hskip -0.5cm
\epsfig{file=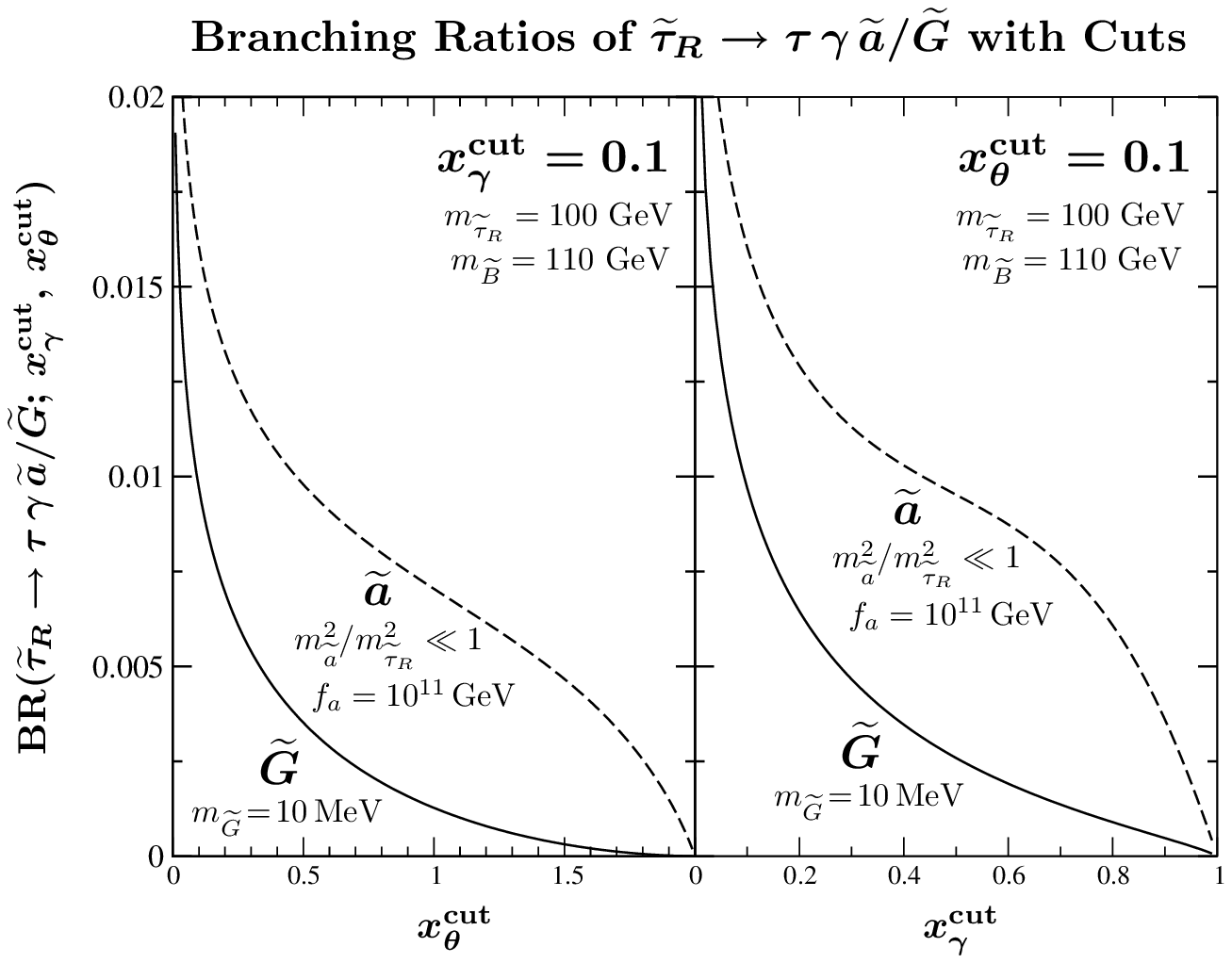,width=14cm}
\caption{The branching ratio of the integrated rate of the
  three-body decay $\stau\to\tau+\gamma+\axino/\gravitino$ with cuts
  as a function of $x_{\theta}^{\mathrm{cut}}$ for $x_{\gamma}^{\rm
    cut}=0.1$ (left) and as a function of $x_{\gamma}^{\mathrm{cut}}$ for
  $x_{\theta}^{\mathrm{cut}}=0.1$ (right). The solid and dashed lines show
  the results for the gravitino LSP and the axino LSP, respectively,
  as obtained with $m_\stau = 100\,\GeV$, $m_\Bino = 110\,\GeV$, $f_a
  = 10^{11}\,\GeV$, $\xi^2 C_{\rm aYY}^2=1$, $m_{\axino}^2/m_{\stau}^2
  \ll 1$, and $\mgravitino = 10\,\MeV$.}
\label{Fig:BranchingRatio}
\efig
this quantity is shown for the gravitino LSP (solid line) and the
axino LSP (dashed line) for $m_\stau = 100\,\GeV$, $m_\Bino =
110\,\GeV$, $f_a = 10^{11}\,\GeV$, $\xi^2 C_{\rm aYY}^2=1$,
$m_{\axino}^2/m_{\stau}^2 \ll 1$, and $\mgravitino =
10\,\MeV$.\footnote{The results shown in Fig.~\ref{Fig:BranchingRatio}
  are basically independent of the Peccei--Quinn scale $f_a$ and the
  gravitino mass $\mgravitino$ provided $\mgravitino/m_\stau\lsim
  0.1$. For larger values of the gravitino mass, the stau NLSP
  lifetime being of $\Order(\mbox{years})$ points already to the
  gravitino LSP scenario as discussed above.}
 In the
left (right) part of the figure we fix $x_{\gamma}^{\mathrm{cut}}=0.1$
($x_{\theta}^{\mathrm{cut}}=0.1$) and vary $x_{\theta}^{\mathrm{cut}}$
($x_{\gamma}^{\mathrm{cut}}$).
The dependence of the branching ratio~(\ref{Eq:BranchingRatio}) on the
cut parameters in the axino LSP case differs qualitatively from the
one in the gravitino LSP case.
Moreover, there is a significant excess of
$BR(\stau_{\mathrm R}\to\tau\,\gamma\, \axino\,; x_{\gamma}^{\mathrm{cut}},x_{\theta}^{\mathrm{cut}})$
over
$BR(\stau_{\mathrm R}\to\tau\,\gamma\, \gravitino\,; x_{\gamma}^{\mathrm{cut}},x_{\theta}^{\mathrm{cut}})$
over large ranges in the cut parameters.
For example, if $10^4$~stau NLSP decays can be analysed and the cuts
are set to $x_\gamma^{\mathrm{cut}}=x_{\theta}^{\mathrm{cut}}=0.1$, we
expect about 165$\pm$13 (stat.) $\stau_{\mathrm
  R}\to\tau\,\gamma\,\axino$ events for the axino LSP and about
100$\pm$10 (stat.) $\stau_{\mathrm R}\to\tau\,\gamma\,\gravitino$
events for the gravitino LSP. Thus, the measurement of the branching
ratio~(\ref{Eq:BranchingRatio}) would allow a distinction to be made
between the axino LSP and the gravitino LSP scenarios. For a smaller
number of analysed stau NLSP decays, this distinction becomes more
difficult. In addition to the statistical errors, details of the
detectors and of the additional massive material needed to stop the
staus and to analyse their decays will be important to judge on the
feasibility of the distinction based on the branching ratios.  We
postpone this study for future work.

\subsection{Differential Distributions in the Three-Body Decays}
\label{Sec:DifferentialDistributions}

Finally, we consider the differential distributions of the visible
decay products in the three-body decays
$\stau\to\tau+\gamma+\axino/\gravitino$ in terms of the quantity
\be
        {1 
        \over 
        \Gamma(\stau_{\mathrm R}\to\tau\,\gamma\, i\,;
        x_{\gamma}^{\mathrm{cut}},x_{\theta}^{\mathrm{cut}})}
        \,\,
        {d^2\Gamma(\stau_{\mathrm R}\to\tau\,\gamma\, i)
        \over
        d x_{\gamma}d \cos\theta}
        \ ,
        \quad 
        i = \axino, \gravitino
        \ ,
\label{Eq:Fingerprint}
\ee
which is independent of the two-body decay, the total NLSP decay
rate, and the Peccei--Quinn/Planck scale. In
Fig.~\ref{Fig:Fingerprint}, 
\befig
\hskip -0.cm
\epsfig{file=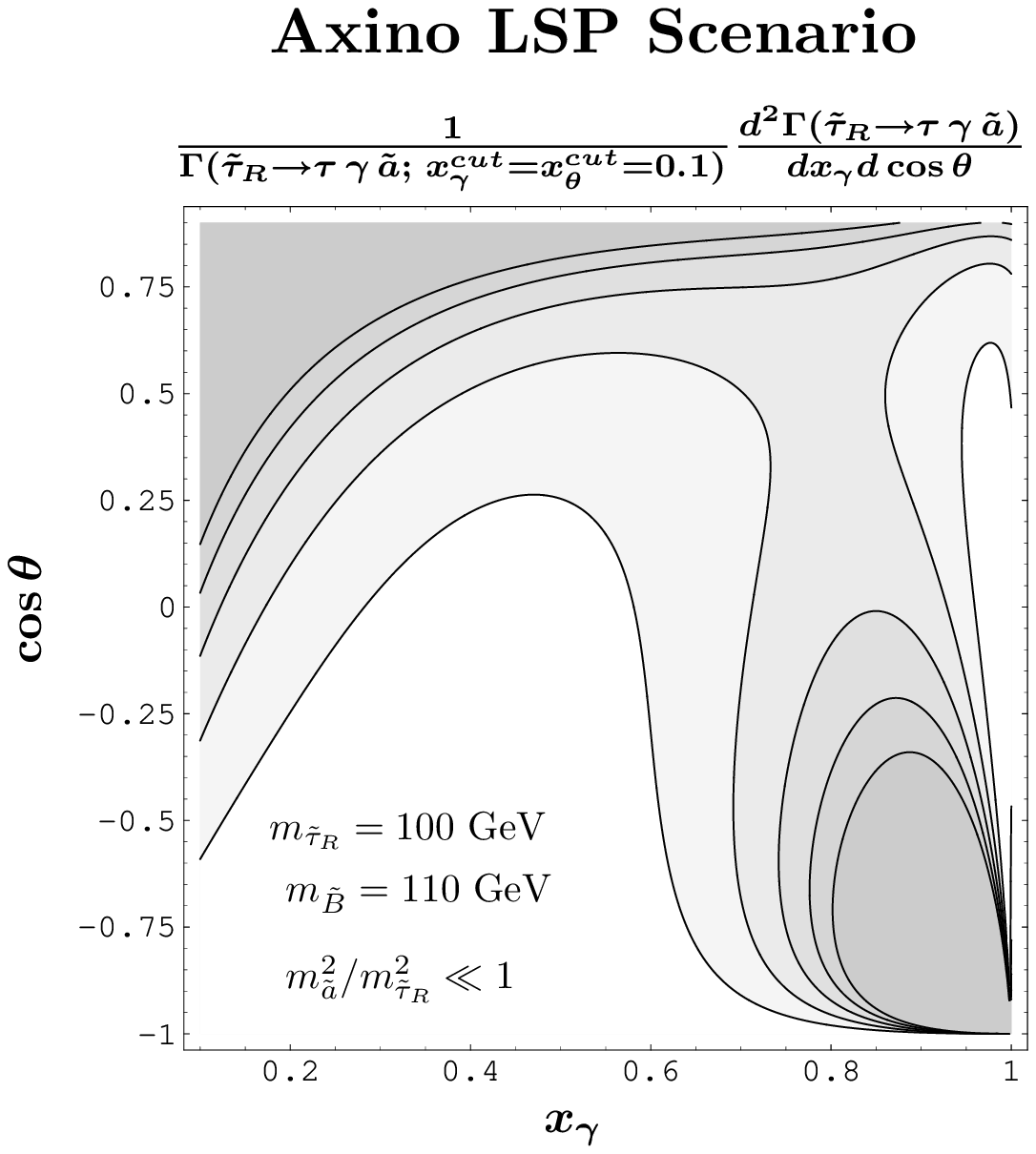,width=8cm}
\hfill
\epsfig{file=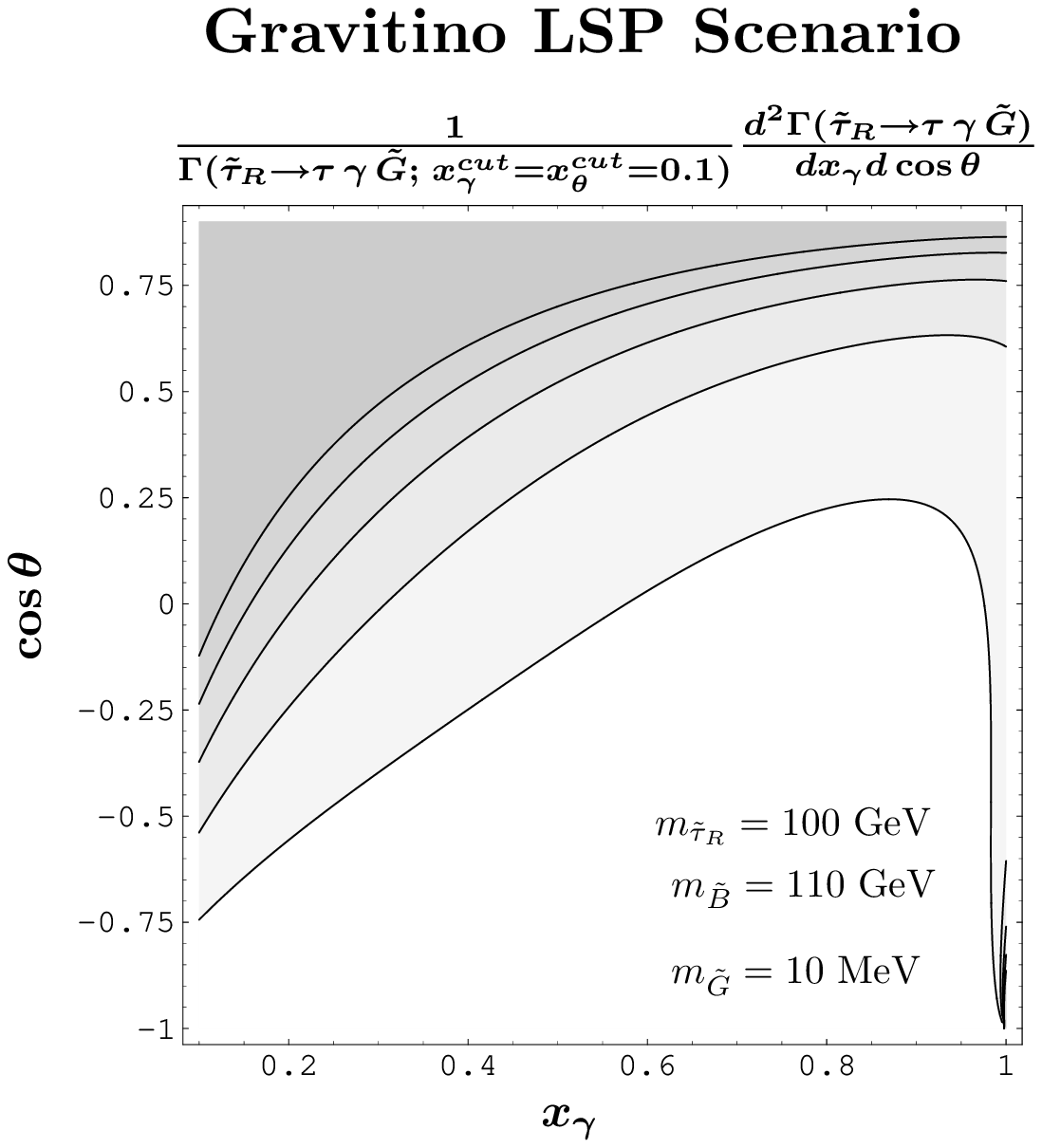,width=8cm}
\caption{
  The normalized differential distributions of the visible decay
  products in the decays $\stau\to\tau+\gamma+\axino/\gravitino$ for
  the axino LSP scenario (left) and the gravitino LSP scenario (right)
  for $m_\stau = 100\,\GeV$, $m_\Bino = 110\,\GeV$,
  $m_{\axino}^2/m_{\stau}^2 \ll 1$, and $\mgravitino = 10\,\MeV$.  The
  cut parameters are set to $x_\gamma^{\mathrm{cut}} =
  x_\theta^{\mathrm{cut}}=0.1$. The contour lines represent the values
  0.2, 0.4, 0.6, 0.8, and 1.0, where the darker shading implies a
  higher number of events.}
\label{Fig:Fingerprint}
\efig
the normalized differential distributions~(\ref{Eq:Fingerprint}) with
$x_\gamma^{\mathrm{cut}} = x_\theta^{\mathrm{cut}}=0.1$ are shown for
the axino LSP scenario (left) and the gravitino LSP scenario (right)
for $m_\stau = 100\,\GeV$, $m_\Bino = 110\,\GeV$,
$m_{\axino}^2/m_{\stau}^2 \ll 1$, and $\mgravitino =
10\,\MeV$.\footnote{A similar comparison between the gravitino and a
  hypothetical spin-1/2 fermion with extremely weak Yukawa couplings
  was performed in Refs.~\cite{Buchmuller:2004rq}.  Note that our
  result for the axino shown in Fig.~\ref{Fig:Fingerprint} differs
  also from the one for the hypothetical spin-1/2 fermion due to
  different couplings.}
In the case of the gravitino LSP, the events are peaked only 
in the region where the photons are soft and the photon and 
the tau are emitted with a small opening angle ($\theta\simeq 0$). 
In contrast, in the axino LSP scenario, the events are also 
peaked in the region where the photon energy is large and 
the photon and the tau are emitted back-to-back ($\theta \simeq \pi$). 
Thus, if the observed number of events peaks in both regions, there is
strong evidence for the axino LSP and against the gravitino LSP. 

To be specific, with $10^4$ analysed stau NLSP decays, we expect about
165$\pm$13 (stat.) events for the axino LSP and about 100$\pm$10
(stat.) events for the gravitino LSP, which will be distributed over
the corresponding ($x_\gamma$,\,$\cos\theta$)-planes shown in
Fig.~\ref{Fig:Fingerprint}. In particular, in the region of
$x_{\gamma}\gtrsim 0.8$ and $\cos\theta \lesssim -0.3$, we expect
about 28\% of the 165$\pm$13 (stat.) events in the axino LSP case and
about 1\% of the 100$\pm$10 (stat.) events in the gravitino LSP case.
These numbers illustrate that $\Order(10^4)$ of analysed stau NLSP
decays could be sufficient for the distinction based on the
differential distributions. To establish the feasibility of this
distinction, a dedicated study taking into account the details of the
detectors and the additional massive material will be crucial, which
we leave for future studies.

Some comments are in order. 
The differences between the two scenarios shown in
Figs.~\ref{Fig:BranchingRatio} and~\ref{Fig:Fingerprint} become
smaller for larger values of $m_\Bino / m_\stau$.  This ratio,
however, remains close to unity for the stau NLSP in unified models.
Furthermore, if $\mgravitino < m_{\axino} <
m_\mathrm{LOSP}$~---~mentioned as case~(iv) in the
Introduction~---~and $\Gamma(\stau \to \axino\,X) \gg \Gamma(\stau \to
\gravitino\,X)$, one would still find the distribution shown in the
left panel of Fig.~\ref{Fig:Fingerprint}.
The axino would then eventually decay into the gravitino LSP and the
axion.  Conversely, the distribution shown in the right panel of
Fig.~\ref{Fig:Fingerprint} would be obtained if $m_{\axino} <
\mgravitino < m_\mathrm{LOSP}$~---~mentioned as case~(iii) in the
Introduction~---~and $\Gamma(\stau \to \axino\,X) \ll \Gamma(\stau \to
\gravitino\,X)$. Then it would be the gravitino that would eventually
decay into the axino LSP and the axion.
Barring these caveats, the signatures shown in
Figs.~\ref{Fig:BranchingRatio} and~\ref{Fig:Fingerprint} will provide
a clear distinction between the axino LSP and the gravitino LSP
scenarios.

\section{Conclusion}
\label{Sec:Conclusion}

Assuming that a charged slepton is the NLSP, we have discussed
signatures of both the gravitino LSP scenario and the axino LSP
scenario in the framework of hadronic, or KSVZ, axion
models~\cite{Kim:1979if+X}. These signatures can be observed at future
colliders if the planned detectors are equipped with 1--10~kt of
additional material to stop and collect charged
NLSPs~\cite{Hamaguchi:2004df,Feng:2004yi}. With calorimetric and
tracking performance, this additional material will serve
simultaneously as a real-time detector, allowing an analysis of the
decays of the trapped NLSPs with high
efficiency~\cite{Hamaguchi:2004df}.

In the scenario in which the axino is the LSP, we have shown that the
NLSP lifetime can be used to estimate the Peccei--Quinn scale $f_a$.
Indeed, if the axino is the LSP, the NLSP decays provide us with a new
way to probe the Peccei--Quinn sector. This method is complementary
to the existing and planned axion search experiments.  The decays of
the NLSP into the axino LSP will also allow us to determine the axino
mass kinematically if it is not much smaller than the mass of the
NLSP. The determination of both the Peccei--Quinn scale $f_a$ and the
axino mass $m_\axino$ will be crucial for insights into the
cosmological relevance of the axino LSP. Once $f_a$ and $m_\axino$ are
known, we will be able to decide if axinos are present as cold dark
matter in our Universe.

In the gravitino LSP scenario, the measurement of the stau NLSP
lifetime can be used to determine the gravitino mass $\mgravitino$
once the mass of the NLSP is known. This will be crucial for insights
into the SUSY breaking mechanism. Moreover, if the gravitino mass can
be determined independently from the kinematics and if the NLSP mass
is known, the NSLP lifetime provides a microscopic measurement of the
Planck scale~\cite{Buchmuller:2004rq}.  Indeed, if the gravitino is
the LSP, the lifetime of the NLSP depends strongly on the Planck scale
and the masses of the NLSP and the gravitino.

We have addressed the question of how to distinguish between the axino
LSP and the gravitino LSP scenarios at colliders. If the mass of the
LSP cannot be measured and if the NLSP lifetime is within the range
$\Order(0.01~{\mbox{sec}})$--$\Order(10~{\mbox{h}})$, we have
found that the NLSP lifetime alone will not allow us to distinguish
clearly between the axino LSP and the gravitino LSP scenarios. The
situation is considerably improved when one considers the three-body
decay of a charged slepton NLSP into the associated charged lepton, a
photon, and the LSP. We have found qualitative and quantitative
differences between the branching ratios of the integrated three-body
decay rate with cuts on the photon energy and the angle between the
lepton and photon directions.
In addition, the differential distributions of the decay products in
the three-body decays provide characteristic fingerprints. 
For a clear distinction between the axino LSP and the gravitino LSP
scenarios based on the three-body decay events, at least of
$\Order(10^4)$ of analysed stau NLSP decays are needed. If the mass of
the stau NLSP is not significantly larger than 100~GeV, this number
could be obtained at both the LHC and the ILC with 1--10~kt of massive
additional material around the main detectors.

\section*{Acknowledgements}

We thank W.~Buchm\"uller, G.~Colangelo, M.~Maniatis, M.~Nojiri,
D.~Rainwater, M.~Ratz, R.~Ruiz de Austri, S.~Schilling, Y.Y.Y.~Wong,
and D.~Wyler, for valuable discussions. We gratefully acknowledge
financial support from the European Network for Theoretical
Astroparticle Physics (ENTApP), member of ILIAS, EC contract number
RII-CT-2004-506222.  This work was completed during an
ENTApP-sponsored Visitor's Program on Dark Matter at CERN, 17~January
-- 4~February 2005. The research of A.B.\ was supported by a
Heisenberg grant of the Deutsche Forschungsgemeinschaft.



\end{document}